# Simulating the Spread of Influenza Pandemic of 2009 Considering International Traffic

Teruhiko Yoneyama
Multidisciplinary Science
Rensselaer Polytechnic Institute
Troy, New York, Unites States
yoneyt@rpi.edu

Mukkai S. Krishnamoothy
Computer Science
Rensselaer Polytechnic Institute
Troy, New York, United States
moorthy@cs.rpi.edu

*Abstract*—**Pandemics have the potential to cause immense disruption and damage to communities and societies. In this paper, we model the Influenza Pandemic of 2009. We propose a hybrid model to determine how the pandemic spreads through the world. The model considers both the SEIR-based model for local areas and the network model for global connection between countries referring to data on international travelers. Our interest is to reproduce the situation using the data of early stage of pandemic and to predict the future transition by extending the simulation cycle. Without considering the tendency of seasonal flu, the simulation does not predict the second peak of the pandemic in the real world. However, considering the seasonal tendency, the simulation result predicts the next peak in winter. Thus we consider the seasonal tendency is an important factor for the spreading of the pandemic.**

## I. INTRODUCTION

Since the spring of 2009, we have experienced influenza pandemic, called Influenza A (H1N1) Pandemic. This pandemic started around March 2009 and it is suspected that it originated in Mexico [1]. On April 24th, WHO announced the emergence of this swine-derived, novel strain of influenza. At that time, some cases had been already confirmed in Mexico and the United States. By May 1, one week after the WHO's alert, cases of outbreak had been reported in other 9 countries. It spread to all over the world in a few months and caused a large number of local infections [2][3].

We expect that the spread of the pandemic is based on the traffic pattern. Thus we propose a hybrid model, which considers both local and global infections. For the local infection, we use the SEIR model considering the each country's condition such as domestic population and population density. For the global infection, we use network based model considering the international travelers which is derived from real data. We compare the simulation result with the real record on the transition of the number of infected cases and find important parameters which influenced the pandemic.

## II. RELATED RESEARCH

Simulating the spreading of infectious disease has been studied in the past. We discuss the differences between this work and other related research. First, a lot of research about simulating disease spread focuses on a prevention/mitigation strategy by comparing the base simulation and an alternative simulation which considers their proposed strategy (e.g. [4][5][6][7][8][9]). In addition, most of existing research simulates with a generated situation which models the real world (e.g. [4][5][8][9][10][11][12]). On the other hand, we focus on the reproduction of the real pandemic using real situation. We model the pandemic, compare the results with real data, and explore the key factors which influenced the spread. Although these critical-factors could provide hints that would help contain the spread of the disease, this paper does not directly propose a prevention strategy.

Second, much research considers the spread of infectious disease from either the local or global point of view (e.g. [6][8][9][11][13]). In addition, much research simulate using one of the equation based (e.g. SIR or SEIR differential equation model), agent based, or network based model (e.g. [11][14][15]). On the other hand, we simulate the pandemic from the global point of view considering local infection in each country. Also, we use a hybrid model which considers both the SEIR based model and network based model using the concept of agent based model.

Third, simulation parameters determine the path of spread. Some research values the basic reproduction number $R_0$ as an influential parameter (e.g. [6][16]). We don't determine $R_0$. In our simulation, we first consider setting the parameters so that the result corresponds with the actual situation in some countries in terms of the number of cases. Then we simulate further experiments using same set of parameters. This is based on the assumption that $R_0$ varies according to country.

## III. Modeling

Previous attempts to model spreading infectious diseases tended to fall into one of two categories. Equation-based models like the SEIR model is suitable for a large-scale spreading of diseases. These models use just a few parameters to reproduce the spreading phenomenon. However it is difficult to reflect detailed situation in countries which have different local infection conditions. Network or agent-based simulation models can theoretically reflect the detail of individual conditions. However, modeling large-scale global diseases is difficult as too many parameters are needed for simulation. Thus we propose a hybrid model. We make a simple model using a small number of parameters and make it capable of simulating a general pandemic.

We simulate using several countries. When we think of an infection in a country, there are three possibilities for new infection; (1) infection from foreign travelers, (2) infection from returning travelers, and (3) infection from local residents. Figure 1 illustrates this concept. We denote the infection-types (1) and (2) as the global infection and the infection-type (3) as the local infection.

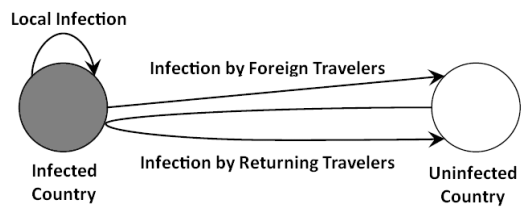

Figure 1: Three Patterns of Infection in a Country

We use the concept of SEIR model which considers four types of agents in each country; Susceptible, Exposed, Infectious, and Removed. Susceptible agents are infected by Infectious agents and become Exposed agents. Exposed agents are in an incubation period. After that period, Exposed agents become Infectious agents. Infectious agents infect Susceptible agents. Infectious agents become Removed agents after the infectious period. Removed agents are never infected again because they are now immune. Figure 2 illustrates this concept of SEIR model.

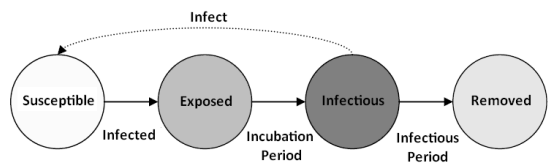

Figure 2: Concept of SEIR

At the beginning of the simulation, the number of Susceptible agents in each country is equal to the population of each country. Then we put an Infectious agent in the origin of the pandemic (i.e. Mexico). The local infection spreads in the origin and the global infection also spreads from the origin to other countries through global traffic. When a country has at least one Infectious agent, that country has the potential for local infection. Figure 3 shows this concept.

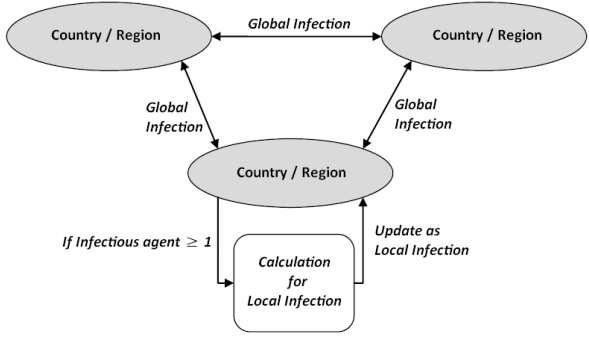

Figure 3: Concept of Simulation Task at One Cycle

The global infection is caused by traffic from infected country. Thus we refer to the number of inbound and outbound traffic. The number of new Exposed agents by the global infection in country $i$ at time $t$, $NEG_i(t)$, is calculated by the expression;

$$NEG_i(t) = I_j(t) \cdot T_{ij} \cdot P_G^*(t) \qquad (1)$$

where $I_j(t)$ is the number of Infectious agents of country $j$ at time $t$. $T_{ij}$ is the total amount of both traffic from country $i$ to $j$ and from $j$ to $i$. $P_G^*(t)$ is the global infection probability at time $t$ and is calculated by the expression;

$$P_G^*(t) = P_G - (D_G \cdot t) \qquad (2)$$

where $P_G$ is the basic global infection probability between countries. $D_G$ is a "deductor" for the global infection. $t$ is time (simulation cycle). $P_G$ and $D_G$ are constants and are uniformly used for every country. Thus the global infection probability $P_G^*(t)$, decreases along the simulation cycle. We assume that, in the real world, the global infection occurs with high probability in early pandemic due to the lack of awareness of the disease. As the disease spreads, people take preventive measures against the infection and the pandemic decreases. We apply this concept in the simulation. The number of Exposed agents in country $i$ at time $t$, $E_i(t)$, is updated by adding $NEG_i(t)$ to $E_i(t)$ at each simulation cycle.

We assume that the local infection probability depends on the population density of a country. Thus if the country is dense, people are more likely to be infected. The basic local infection probability of country $i$, $P_{Li}$ is given by the expression;

$$P_{Li} = Density_i \cdot C_1 + C_2 \qquad (3)$$

where $Density_i$ is population density of country $i$, obtained by real data. Thus $Density_i$ differs in country. $C_1$ and $C_2$ are constants and are used for simulation in every country.

We assume that the number of new Exposed cases of a country by the local infection depends on the number of Susceptible agents and the number of Infectious agents at that time. Thus the number of new Exposed agents by the local infection in country $i$ at time $t$, $NEL_i(t)$, is calculated by the expression;

$$NEL_i(t) = S_i(t) \cdot I_i(t) \cdot P_{Li}^*(t) \qquad (4)$$

where $S_i(t)$ us the number of Susceptible agents of country $i$ at time $t$. $I_i(t)$ is the number of Infectious agents of country $i$ at time $t$. $P_{Li}^*(t)$ is the local infection probability at time $t$ and is calculated by the expression;

$$P_{Li}^*(t) = P_{Li} - (D_L \cdot t) \qquad (5)$$

where $P_{Li}$ is the basic local infection probability of country $i$ which is obtained by equation (3). $D_L$ is a "deductor" for the local infection and is a constant which is used for every country. $t$ is time (simulation cycle). Similar to the global infection, the local infection probability $P_{Li}^*(t)$ decreases as the simulation cycle increases. This reflects people's awareness. The number of Exposed agents in country $i$ at time $t$, $E_i(t)$, is updated by adding $NEL_i(t)$ to $E_i(t)$ at each simulation cycle.

Table 1 summarizes parameters in the simulation. We have eight controllable parameters which are denoted as constants in Table 1. These parameters are used for every country uniformly. Other parameters are derived from real data and depend on country.

Table 1: Parameters in Simulation

| Parameter | Description | Attribution | |
|---|---|---|---|
| | | (a)Global or (b)Local | (1) Constant or (2) Depend on Country |
| $P_G$ | Global Infection Probability | (G) | (1) |
| $P_{Li}$ | Local Infection Probability of County $i$ | (L) | (2) |
| $D_G$ | Deductor for Global Infection Probability | (G) | (1) |
| $D_L$ | Deductor for Local Infection Probability | (L) | (1) |
| $C_1$ | Constant for Local Infection Probability | (L) | (1) |
| $C_2$ | Constant for Local Infection Probability | (L) | (1) |
| *Incubation_Period* | Incubation Period | (G) and (L) | (1) |
| *Infectious_Period* | Infectious Period | (G) and (L) | (1) |
| *Run_Cycle* | Run Cycle of Simulation | (G) and (L) | (1) |
| *Density$_i$* | Actual Population Density of Country $i$ | (L) | (2) |
| *Population$_i$* | Actual Population of Country $i$ | (L) | (2) |
| $T_{ij}$ | Amount of Traffic between Country $i$ and $j$ | (G) | (2) |

## IV. STATISTICS ON NUMBER OF CASES

We need real data on the number of cases to use it in our simulation and to compare it with our simulation result. However, After July 6th 2009, WHO has been publishing the number of expected cases by regions instead of the number of laboratory confirmed cases in each country [3]. This is because it is difficult to count the exact number of laboratory confirmed cases in a country due to the large number of patients suspected to have the illness. Thus it is difficult to obtain the data on the exact number of infected cases.

In order to observe the transition of the pandemic with a fixed criterion, we refer to the percentage of visits for ILI (Influenza-like Illness). This method is used as the criterion for the epidemic in the United States, Canada, and European countries. Although this method considers not only Influenza A/H1N1 but also other types of influenza, this can be a good criterion to observe the transition of the pandemic. Figure 4 shows the transition of weekly percentage of visits for ILI form week 16 of 2009, the beginning of the pandemic, to week 5 of 2010 [17][18][19][20][21][22]. We divide into two figures due to the different scale. Figure 4 (a) shows the transition in the United States and Canada. Figure 4 (b) shows the transition in 8 European countries. The figures show the tendency that each country had the first peak in early or middle summer. Although once a depression came after the first peak, the second peak arrived during the winter season again with much larger cases in many countries.

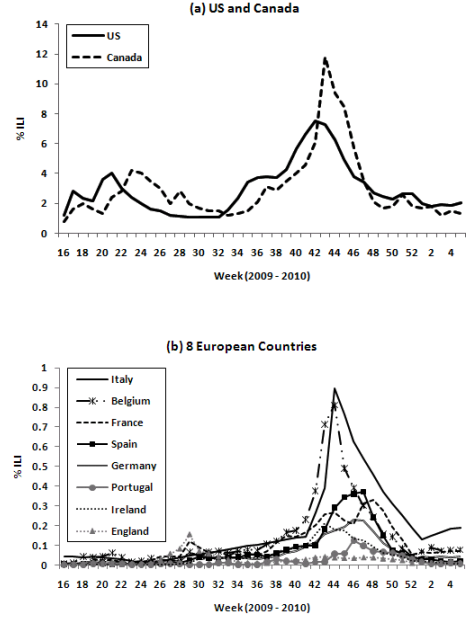

Figure 4: Transition of Percentage of Visits for ILI from Week 16 2009 through Week 5 2010 ((a) US and Canada, (b) 8 European Countries, Created based on [17][18][19][20][21][22])

## V. SIMULATION AND RESULATS

For the global infection, we refer to the number of travelers between countries referring to [23][24]. Since we assume that the origin of the pandemic is Mexico, we look at the number of travelers from/to Mexico. We sum up the number of inbound and outbound travelers in Mexico, and find some countries which have strong relationship with Mexico in terms of the number of travelers. From this, we select Top 5 countries as the United States, Canada, France, Spain, and United Kingdom. We assume that a country which has strong relationship with Mexico is likely to import an early case of the influenza. According to WHO [3], all of these 5 countries were infected within 8 days from the WHO's first announcement of the emergence of the novel influenza on April 24th. The United States had been infected before that.

Thus these 5 countries have higher possibility to be infected directly from Mexico. Next we examine the number of travelers from/to these 5 countries and find Top 5 related countries of these 5 countries. By this way, we find a total of 13 countries including Mexico; Belgium, Canada, China, France, Germany, Ireland, Italy, Japan, Mexico, Portugal, Spain, United Kingdom, and the United States. Of these 13 countries, we simulate 11 countries excluding China and Japan since these countries use different counting methods but percentage of visits for ILI. We complete the travelers table among these 11 countries and use it for our simulation. For the local infection, we refer to the actual population and population density of these 11 countries referring to [25].

At the beginning of the simulation, we place 18 Infectious agents in Mexico and 7 Infectious agents in the United States, based on the WHO's report as of April 24th 2009 [3]. Our interest is to predict the future transition of the spread with using data of the early period of the pandemic. Thus at first we set the parameter so that the simulation result corresponds to the intermediate situation as of July 6th 2009, the date of the last report for each country by WHO. Then we extend the simulation run cycle with using same parameters in order to simulate the future. We set the parameter values so that the number of cumulative cases in our simulation result becomes close to the number of the laboratory-confirmed cases in 3 countries as of July 6th, the United States, Mexico, and Canada, whose number of reported cases are most significant among all countries at that time. We simulate for 10 weeks from April 24th. For this simulation, we set the run cycle as 70 so that 7 cycles in simulation correspond to 1 week in the real world.

Figure 5 shows the comparison between our simulation result and the real data as of July 6th 2009 in 11 countries. The number of cases in the United States, Mexico, and Canada almost corresponds to each other.

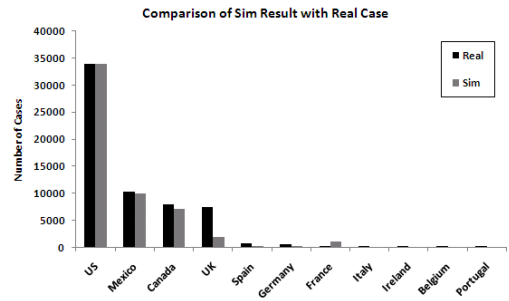

Figure 5: Comparison of Cumulative Cases as of July 6th between Simulation Result and Real Data in 11 Countries

Then we extend the simulation cycle to simulate the situation after July 6th. We extend to 364 cycles, which matches 52 weeks, one year, in the real world. Figure 6 shows the simulation result regarding the expected infection route. This figure shows how the pandemic spread from Mexico to world by our simulation result. We assume that the United States is infected by Mexico at a very early period of the pandemic. Next, Canada, United Kingdom, and France are also infected through the United States. Then, the pandemic spreads from France to its neighboring countries such as Belgium, Germany, and Italy. The pandemic which spread to United Kingdom reaches Spain and Ireland. Spain also infects Portugal.

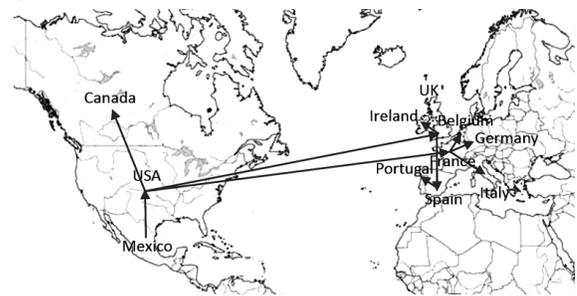

Figure 6: Comparison of Cumulative Cases as of July 6th between Simulation Result and Real Data in 11 Countries

According to our simulation, the pandemic which occurs in Mexico tends to spread to North America and European countries through the United States. In Europe, France and United Kingdom act as hub countries for the spread. Considered the number of travelers in the real data, the pandemic originates from Mexico tends to spread to other countries through North America or Europe. South America is geographically near to Mexico, but it is not infected earlier than other regions since it is not closer to Mexico compared with other regions in terms of traffic. Thus South America tends to be infected after European countries are infected.

Next we look at the simulation result on the transition of infected cases. Figure 7 shows the transition of number of infected cases as of Cycle 364.

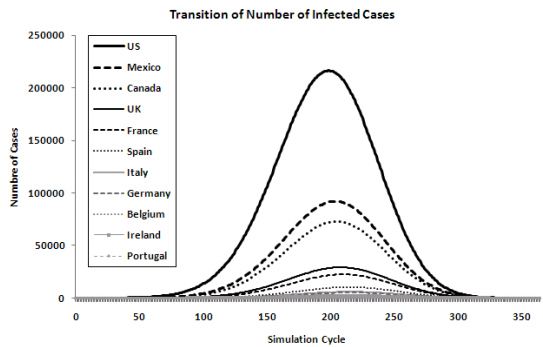

Figure 7: Comparison of Cumulative Cases as of July 6th between Simulation Result and Real Data in 11 Countries

We have simulated the pandemic assuming it is transient; once each country has its peak, the number of new cases decreases. However, in the real world, there have been two peaks in 11 countries. Figure 4 shows that each country has the first peak in early or middle summer of 2009 and has the second peak during the winter season. Our simulation results do not reproduce the fluctuation of the spread and the two peaks. We expect that the pandemic is much influenced by the seasonal conditions.

In order to realize the seasonal factors, we consider the historical tendency of influenza in each country. Thus we refer to the weekly percentage of visits for ILI in past few years. We expect that the percentage of visits for ILI is influenced by seasonal conditions and comprehensively indicates the spread of influenza. Figure 8 shows the transition of the average weekly percentage of visits for ILI in 10 countries in historical data. Figure 8 (a) shows the United States and Canada [21][26]. Figure 8 (b) shows 8 European countries [19][22].

We refer to data in 1999 through 2007 for the United States, data in 1996 through 2008 for Canada, and data in 2004 through 2008 for 8 European countries, and take the average for each week. For the convenience, we show the transition from 16th week of year since the pandemic starts from that week.

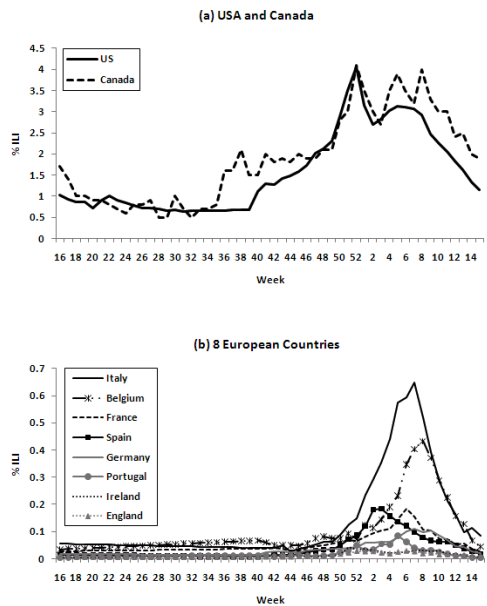

Figure 8: Transition of Average Weekly Percentage of Visits for ILI in Historical Data ((a) US and Canada, (b) 8 European Countries, Created based on [19][21][22][26])

In order to consider this data in our simulation, we apply the weekly percentage of visits for ILIU for the local infection. Let $ILI_i(t)$ be the percentage of visits for ILI of country $i$ at time $t$. Then the number of new Exposed agents of country $i$ by the local infection at time $t$, $NEL_i(t)$, is calculated by the expression;

$$NEL_i(t) = S_i(t) \cdot I_i(t) \cdot P_{Li}^*(t) \cdot ILI_i(t) \qquad (6)$$

$S_i(t)$ is the number of Susceptible agents of country $i$ at time $t$ and $I_i(t)$ is the number of Infectious agents of country $i$ at time $t$. $P_{Li}^*(t)$ is the local infection probability of country $i$ at time $t$ which is considered the population density and the deduction for the local infection.

Thus $NEL_i(t)$ fluctuates depending on not only the number of Susceptible and Infectious agents and the local infection probability which depends on the population density, but also the average weekly percentage of visits for ILI. Since the average percentage of visits for ILI in past few years comprehensively indicates the influence by seasonal condition of a country, we expect that it can be used for the local infection. Since we regard 7 cycles as 1 week in our simulation, we apply the percentage of visits for ILI in a country as follows. The simulation starts from week 16 of 2009. Thus, for each country for the first 7 cycles in our simulation, we apply the average percentage of visits for week 16 in the historical data in a country as the local infection for the country. For next 7 cycles, we apply the percentage for week 17 in the historical data.

Then we simulate again considering the seasonal factor. As well as the previous experiment, we begin with reproducing the early situation, the situation as of July 6[th] in the United States and Canada. Since it is expected that the early situation in the real world was also influenced by the seasonal factor, we change the parameters a little so that the numbers of cumulative cases in the simulation result correspond with that in the real data as of July 6[th].

Then we extend the simulation cycle to 364 to predict the situation for one year. Our interest is to predict the transition of pandemic by using only data at the early time of the pandemic. Since the data on the percentage of visits for ILI are available in the United States, Canada, and 8 European countries, we simulate with these 10 countries in the following sections. In Figure 9, we show the simulation result which takes into account the historical seasonal flu data. In Figure 9 (a), we show the transition of the total number of ILI in the United States and Canada. Although all ILI cases are not by the Influenza A/H1N1, this data is useful to find the tendency of the transition of new infected cases by the novel influenza as well as Figure 4 since most cases are by A/H1N1 in this year [20][21]. Note that in Figure 9 (b), the number of cases in the simulation result is based on the number of laboratory-confirmed cases. Thus it is expected that the actual number of infected cases is much more since many infected person are not confirmed at a laboratory. Also, since we apply the percentage of visits for ILI in each country for the local infection, it may not be good to compare the simulation results with the real data in terms of the number. Thus we focus on the transition of the spread rather than the number of cases.

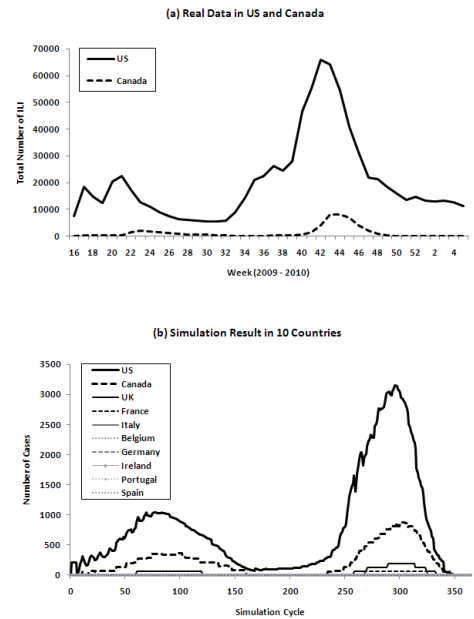

Figure 9: Transition of Number of New Infected Cases in Simulation with Considering Historical Seasonal Flu Data ((a) Real Data in US and Canada for Week 16 2009 – Week 5 2010, Created based on [17][18][20][21], (b) Simulation Result in 10 Countries as of Cycle 364)

By applying the historical percentage of visits for ILI, simulation result can reproduce the two peaks of the pandemic in summer and winter seasons. The United States and Canada have their first peak around Cycle 80. After some decreases, the number of cases increases from around Cycle 230 due to the winter season. The second peak comes around Cycle 300. After the peak, the number of cases decreases. The tendency of the transition is similar to that in the United States and Canada in the real world; the first spread is in the mid-summer of 2009 and the second spread is after fall of 2009. For European countries, although there is no data available on the total number of ILI, Figure 4 can be comparable to compare the tendency of the transition.

In Figure 9, the transition in the simulation result seems a little delay compared with that in the real data in Figure 4. This is because the simulation refers to the historical tendency. As Figure 8 shows, in many countries, the peak of seasonal flu usually comes around week 52 through week 8 of the next year. Thus the simulation result reflects this tendency. On the other hand, the peak of winter season in 2009-2010 came week 42 through week 48 in many countries, as Figure 4 shows. Therefore the peak of influenza of 2009-2010 is earlier than that of usual year. One possible reason for the earlier peak may be people's awareness. People were aware of the spreading of the novel influenza from summer, and that made the quick countermeasure against the second peak, which resulted in the earlier convergence of the peak.

## VI. Conclusions

In this paper, we simulated the Influenza Pandemic of 2009, considering the international travel. Our interest was to predict the future transition of the spread with using data of the early pandemic. At first, we reproduced the situation of the early stage of the pandemic. Then we extended the simulation cycle to predict the future transition. However, our model didn't consider the seasonal conditions. In the Northern hemisphere countries, the second wave was coming in winter season of 2009-2010. To rectify this, we took into account historical data for the seasonal flu and simulated again. The modified simulation result showed two peaks which came in the middle through late winter in 2010. Thus we conclude that this pandemic is much influenced by the seasonal flu's tendency. We used only the data at the early pandemic, as of July 6[th] 2009, in our simulation and we found that our simulation result is almost identical tendencies by comparing with the real situation.